\documentclass[10pt,letterpaper,twocolumn]{article} 

\usepackage{ol}

\newcommand{\ket}[1]{\left| #1\right\rangle}

\newcommand{\ketbra}[2]{\left| #1\right\rangle\!\left\langle#2\right|}

\begin{document}
\twocolumn[

\title{On decoherence of electromagnetically-induced transparency in atomic vapor}

\author{E. Figueroa, F. Vewinger, J. Appel, A. I. Lvovsky}

\address{Institute for Quantum Information Science, University of
  Calgary, Alberta T2N 1N4, Canada }

\date{\today}


\begin{abstract}
We report characterization of electromagnetically induced transparency (EIT) resonances in the D1
line of $^{87}$Rb under various experimental conditions. The dependence of the EIT linewidth on the
power of the pump field was investigated, at various temperatures, for the ground states of the
lambda-system associated with different hyperfine levels of the atomic 5S$_{1/2}$ state as well as
magnetic sublevels of the same hyperfine level. Strictly linear behavior was observed in all cases.
A theoretical analysis of our results shows that dephasing in the ground state is the main source
of decoherence, with population exchange playing a minor role.
\end{abstract}

]


{\bf Introduction.} Since the discovery of electromagnetically-induced transparency (EIT) in the
early nineties \cite{harris97:_elect}, this phenomenon has received significant attention due to
its vast range of applications, in particular in quantum computation and quantum communication
\cite{Duan20015,Fleisch_77}. They include storage of quantum states of light
\cite{lukin01,kuzmichnature,lukinnature}, logic gates \cite{SchmidtImamoglu}, magnetometry
\cite{magnetometry}, and routing of optical information \cite{RATOS}. Many of these applications
require thorough understanding of the phenomena responsible for the width of EIT resonances, in
particular of the relevant decoherence processes.

Decoherence in EIT is caused by several mechanisms, such as flight-through broadening, population
exchange, atom-atom and atom-wall collisions, etc., but it is still not clear which one is the most
significant. Insight into this question can be gained by measuring the width of the EIT resonance
as a function of the pump field intensity. Most existing experiments in atomic vapors
\cite{lin1,lin2,lin3,Nov06} showed this dependence to be linear, with an exception of the work by
Ye and Zibrov~\cite{Ye} (which was performed in unusual conditions, without buffer gas and with a
very small beam diameter). On the other hand, an existing theoretical treatment \cite{Javan1,Lee},
assuming the population exchange between the ground levels $\ket{b}$ and $\ket{c}$ [Fig.~1(a)] to
be the main source of decoherence, predicts a non-linear dependence for weak pump powers.

In the present work, we measured the width of the EIT resonances on the D1 transition in rubidium
vapor in a variety of settings. Our results show a consistent linear behavior with the $y$-axis
intercepts on the order of a few kilohertz. Based on these results, we propose an alternative
theory based on pure dephasing (i.e.~decay of the off-diagonal density matrix elements) as the
dominant decoherence mechanism. This treatment does predict linear dependencies, yielding much
better fits to our data. From the fits, we also obtain the ground states' decoherence rates on the
scale of a few kHz, which are consistent with independent verifications.




\begin{figure}[b]
    \begin{center}
      \includegraphics[width=0.45\textwidth]{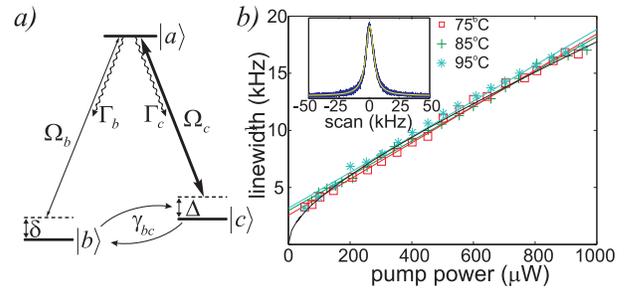}
        \caption{\small{(a) The relevant atomic level structure. The excited state
    $\ket{a}$ is coupled to two ground states $\ket{b}$, $\ket{c}$ by a weak signal field with Rabi
    frequency~$\Omega_b$ and detuning $\delta$ and a stronger driving
    field with Rabi frequency~$\Omega_c$ and detuning~$\Delta$.
    $\Gamma_b$ and $\Gamma_c$ are rates of spontaneous emission into
    the respective states $\ket{b}$ and $\ket{c}$ whereas $\gamma_{bc}$
    describes the decay rate of coherences between these states. (b) Measured width of the Zeeman configuration EIT resonances as a function
        of the pump laser power for different temperatures
        together with linear fits and a fit to the theory
        of Javan {\it et al.}\cite{Javan1,Lee}, assuming a decoherence rate of
        $\gamma_{bc}=$110~Hz.
        The inset shows an example of the measured EIT resonance.}}
        \label{FWHM2}
    \end{center}
\end{figure}

{\bf Experimental setup.} The experiments were performed in atomic $^{87}$Rb vapor at temperatures
60--100$^{\circ}$C, which correspond to the optically thick regime, using two $\Lambda$ energy
level configurations. The first one was formed by one of the hyperfine sublevels of the $5P_{1/2}$
state and two hyperfine sublevels of the $5S_{1/2}$ state. In the second one, the ground states
were represented by different magnetic sublevels of the same hyperfine level. We refer to these
settings as ``hyperfine'' and ``Zeeman'', respectively.

In the Zeeman setup, the pump and signal fields of wavelength $\lambda=795$ nm couple pairs of
Zeeman sublevels of the atomic ground state ($5S_{1/2}$, $F=2$) via the excited state ($5P_{1/2}$,
$F'=2$). The light source is a Coherent MBR-110 Ti:Saphire laser with a narrow spectral width
($\sim$ 100kHz) and high long-term stability. Relative frequencies of both fields were precisely
controlled by acousto-optical modulators. After modulation, both fields were recombined in a
polarizing beam splitter, converted to orthogonal, circular polarizations by means of a quarter
wave plate and directed into a 5-cm long rubidium vapor cell with 1 Torr neon buffer gas, located
inside a magnetically isolated oven. The measured diameter of the beam just before the cell was
$\sim$ 10 mm. After the cell, the linear polarization was recovered and the signal field detected.


In the hyperfine configuration, the signal field was obtained from an additional diode laser that
was phase locked, at 6.8 GHz, to the Ti:Sapphire laser to ensure a two-photon resonance with the
($\ket{b}=\ket{5S_{1/2},F=1},\ket{a}=\ket{5P_{1/2},F=2},\ket{c}=\ket{5S_{1/2},F=2})$ transition. In
this setting, we used linear polarizations for the pump and signal.

To measure the full width half maximum (FWHM) of the EIT resonance, we swept the signal field
frequency. A typical scan is depicted in the inset of Fig.~1(b) and approximates a Lorentzian
distribution. The pump power was varied from 100 $\mu$W to 1.2 mW ($\Omega_{c} \sim$ 1-3 MHz) while
the signal power was kept constant at about 20 $\mu$W ($\Omega_{b} \sim$ 500 kHz).

Fig.~1(b) shows the results of the Zeeman measurement. The behavior is linear, temperature
independent, and shows a $y$-intercept of 3 kHz. Similar measurements were performed in a 10-Torr
Ne buffer gas cell (not depicted), showing a similar behavior and a slightly different
$y$-intercept.

Experimental results for the hyperfine configurations are presented in Fig.~2. Similarly, a linear
behavior is observed but the slope and the zero crossing do depend on temperature. Additional
measurements were done with a 0.1-Torr cell (not depicted), showing a similar trend.

Both figures also display the best fit obtained using the theory of Javan {\it et
al.}\cite{Javan1,Lee}. These fits do not follow well the experimentally observed data and yield
unrealistically low values for the decoherence rates: 110 Hz and 117 Hz, respectively. We conclude
that the population exchange cannot be the dominant decoherence mechanism in our cells.

\begin{figure}[b]
    \begin{center}
      \includegraphics[width=0.45\textwidth]{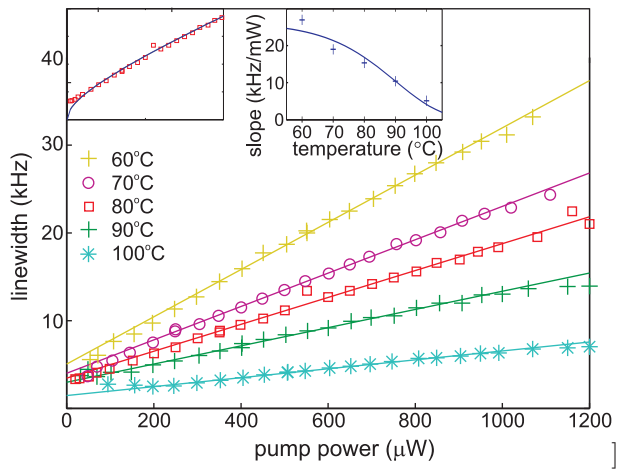}]
        \caption{\small{Dependence of the EIT linewidth on the pump power for the hyperfine configuration. The lines are linear fits to the measured data.
        The first inset shows a fit of the population exchange theory \cite{Javan1} to the data acquired at T=80$^{\circ}$C, with
        $\gamma_{bc}=$117~Hz. The second inset shows the temperature dependence of the experimentally determined slopes
         together with a theoretical estimate.}}
        \label{EitWidth_Hyp}
    \end{center}
\end{figure}



{\bf Theory.} Motivated by this discrepancy, we developed an alternative theory assuming that the
ground state decoherence is dominated by dephasing, i.e. decay of the off-diagonal density matrix
element $\rho_{bc}$ at a rate $\gamma_{bc}$. With this assumption, and the notation defined in
Fig.~1(a), we first determine the stationary density matrix of a single atom. Without the signal
field, it is just $\rho^{(0)}=\ketbra{b}{b}$: all atoms are pumped into the $\ket{b}$ state. In the
first (linear) order with respect to the signal, one obtains a well-known expression
\cite{Scullysbook}
\begin{equation}
  \label{eq:3}
  \rho_{ab}^{(1)} = \Omega_b \, \frac{1}{\Delta-\delta_2
  + \frac{|\Omega_c|^2}{i\gamma_{bc}+\delta_2} - i \frac{\Gamma}{2}},
\end{equation}
where $\Gamma=\Gamma_b+\Gamma_c$ is the inverse lifetime of the excited state and $\delta_2 =
\Delta-\delta$ is the two-photon detuning.


Atoms in a hot gas atoms ``see'' the light fields Doppler shifted according to their momentary
momentum. In our experiment, both fields are co-propagating, and because their optical frequencies
are similar, they experience a similar Doppler shift: the two-photon detuning $\delta_2$ is almost
independent of the individual atoms' motion while the pump field detuning $\Delta$ is not.

Assuming that the pump field is tuned to the center of the Doppler-broadened line, averaging
expression (\ref{eq:3}) over all atoms weighted by their velocity distribution $p(\Delta)=p(2\pi
v/\lambda)$ gives the susceptibility for the signal field
\begin{equation}
  \label{eq:4}
  \chi_b = \frac{\wp}{\Omega_b} \int p(\Delta) \, \rho_{ab}
  \,d\Delta,
  \qquad {\rm with } \wp=\frac{N \mathcal{D}_{ab}^2}{\hbar
  \varepsilon_0},
\end{equation}
where $N$ is the atomic density, $\mathcal{D}_{ab}$ the dipole moment of the
$\ket{a}\leftrightarrow\ket{b}$ transition, $\varepsilon_0$ is the free space permittivity and
\begin{equation}
  \label{eq:5}
  p(\Delta)=\frac{\sqrt{\ln 2}}{W_d\sqrt{\pi}} e^{-\ln 2\left(\frac{\Delta}{W_d}\right)^2},
\end{equation}
with $2W_d=(4\pi/\lambda)\sqrt{2\ln 2 k_B T/m_{Rb}}\approx 0.55$ GHz being the FWHM of the
Doppler-broadened line.

 To simplify our calculations, we approximate the Gaussian Boltzmann distribution
by a Lorentzian of the same width and maximum:
\begin{equation}
  \label{eq:6}
  p_{\rm Lorentz}(\Delta)=\frac{\sqrt{\ln 2}}{W_d\sqrt{\pi}}\
  \frac{1}{1+\left(\frac{\Delta}{W_d}\right)^2}\,.
\end{equation}
Eq. (\ref{eq:6}) approximates the Gaussian distribution reasonably well for small $\Delta$; it
differs only in its wings, which correspond to high-velocity atoms that see both fields far detuned
and thus interact only weakly. Performing the integration (\ref{eq:4}) in this approximation, we
find the average susceptibility for the $\Omega_b$ field:
\begin{equation}
  \label{eq:7}
  \chi_b = 2 \wp \sqrt{\pi \ln 2}
  \frac{i \gamma_{bc} + \delta_2}{(\gamma_{bc}- i \delta_2)(\Gamma+2W_d-2 i \delta_2)+2 |\Omega_c|^2}.
\end{equation}

The EIT linewidth is several orders of magnitude smaller than the Doppler width $2W_d$, hence we
can drop the $2 i \delta_2$ term in the denominator. The (intensity) absorption coefficient
$\alpha$ is then of Lorentzian form
\begin{equation}
  \label{eq:8}
  \alpha(\delta_2) = \frac{\omega}{c} \textrm{Im}(\chi_b)
  = \alpha_{\rm max} - \frac{\alpha_{\rm max}-\alpha_{\rm min}}{1+\left(\frac{2 \delta_2}{\mathrm{FWHM}}\right)^2}
\end{equation}
with
\begin{eqnarray}
  \alpha_{\rm max} &  = & 2\frac{\omega \wp}{c}\sqrt{\pi \ln 2}  \frac{1}{2W_d+\Gamma}\, , \\
  \alpha_{\rm min} &  = & 2\frac{\omega \wp}{c}\sqrt{\pi \ln 2}  \frac{1}{2W_d+\Gamma+2 \frac{|\Omega_c|^2}{\gamma_{bc}}}, \\
  \mathrm{FWHM} & = & 2 \gamma_{bc} + \frac{4 |\Omega_c|^2}{2W_d +\Gamma}. \label{eq:9}
\end{eqnarray}

Since $|\Omega_c|^2$ is proportional to the beam intensity, the EIT linewidth scales linearly with
the pump power and intersects the $y$-axis at a minimum of $2 \gamma_{bc}$.

{\bf Discussion.} As evident from Figs.~1 and 2, linear dependence (\ref{eq:9}) provides an
excellent fit to our experimental data. The $y$-axis intercept is twice the ground state
decoherence rate, which can be measured independently, e.g., by means of light
storage\cite{lukin01}. With our setup, storage times of 100-250 $\mu$s were observed, which is in
reasonable agreement with the measured intercepts.

We note that the population exchange theory \cite{Javan1,Lee} predicts that at high pump intensity,
the dependence of the EIT linewidth on the pump power approaches linear,
\begin{equation}
\mathrm{FWHM} \to 4 \gamma_{pe}\frac{W_d}{\Gamma} + \frac{2 |\Omega_c|^2}{W_d},
\end{equation}
$\gamma_{pe}$ being the population exchange rate. This asymptotic dependence has the slope similar
to that in Eq. (\ref{eq:9}), but its intercept, for $\gamma_{pe}\sim\gamma_{bc}$, is higher by a
factor of $\sim 180$. Therefore, even a minute population exchange rate would have a significant
effect on the dependencies studied. Our experimental results thus show that the fraction of
population exchange mechanism in the ground state decoherence is indeed minor. We can estimate
$\gamma_{pe}$ to be below 50 Hz.

Decoherence in atomic vapor cells is known to be dominated by the flight-through mechanism. The
atoms arrive into the interaction area in an arbitrary ground state, so one would expect at least
some signature of population exchange decoherence, and it is surprising that none is present. We
explain this by the geometry of the laser beams, which was close to Gaussian. Before entering the
interaction area, an atom, initially in a random ground state, propagates through the ``wings'' of
the Gaussian profile. In this region, the signal field is negligible, but the pump field is already
sufficiently strong to pump the atom out of $\ket{c}$. When entering the central part of the
interaction region, most of the atoms will be in state $\ket{b}$, albeit with a random phase.

The temperature dependence of the slope in the hyperfine configuration can be explained by
absorption of the pump laser by the saturated rubidium vapor. We quantified this effect by
integrating Eq.~(\ref{eq:7}) numerically over the length of the cell, assuming that the pump
absorption is governed by Beer's law. The temperature dependence of the slope derived by this
procedure (Fig.~2, second inset) was compared to the measured slopes. The best fit was obtained for
the 12-mm beam diameter, which agrees well with our measured diameter of 10~mm. In the Zeeman
configuration, the pump absorption was negligible at all temperatures, which explains a constant
slope.


{\bf Conclusions.} We have experimentally observed the behavior of the width of an EIT resonance as
a function of the pump power in different $\Lambda$ configurations and temperatures. Consistently
linear behavior was observed and explained theoretically with an assumption that the main ground
state decoherence mechanism is dephasing.

{\bf Acknowledgments.} This work has been sponsored by NSERC, CFI, AIF, and CIAR. EF also thanks
the DAAD for its sponsorship. AL's email is lvov@ucalgary.ca


\end{document}